\documentclass[twocolumn,english]{article}
\usepackage{bookman}
\usepackage[T1]{fontenc}
\usepackage[latin1]{inputenc}
\usepackage{graphicx}

\makeatletter

\usepackage{babel}
\makeatother
\begin{document}
\begin{minipage}[c]{1.0\textwidth}%
\begin{center}\textbf{\Large Relating RHIC and SPS Heavy Ion Data}\end{center}{\Large \par}

\noindent \begin{center}\textbf{\large Klaus WERNER }\end{center}{\large \par}

\noindent \begin{center}\textit{SUBATECH, Université de Nantes --
IN2P3/CNRS-- EMN,  Nantes, France}\emph{}\\
\emph{werner@subatech.in2p3.fr}\end{center}

Abstract: We have demonstrated recently that RHIC heavy ion data concerning
particle production at small and intermediate transverse momenta become
very transparent, if one realizes that there are two sources of particles
production: the corona -- due to the interactions of the peripheral
nucleons of either nucleus, and the core -- representing the high
density central zone. We extend our analysis to SPS heavy ion data.
Again, we find that the nontrivial centrality dependence of particle
production (being stronger than at RHIC) is simply due to an increasing
corona contribution with decreasing centrality. The core contribution
shows no centrality dependence (concerning particle ratios), and can
be described in exactly the same way as at RHIC, with one exception:
there is somewhat less collective radial flow at SPS compared to RHIC.
Apart of this (and a trivial volume effect), the core portions produced
in different colliding systems at different energies (RHIC, SPS) look
the same.\end{minipage}%

\bigskip{}

One of the spectacular results from SPS heavy ion research is the
very strong enhancement of strange (in particular multi-strange) baryons
compared to proton-proton scattering. Although this goes qualitatively
into the right direction, having in mind the production of a quark-gluon-plasma
\cite{rafelski}, a quantitative understanding is still lacking.

We do not really claim to explain these data, we simply want to stress
that they show an almost trivial behavior, if one takes into account
the {}``corona effect'': the peripheral nucleons of either nucleus
essentially perform independent pp or pA-like interactions, with a
very different particle production compared to the high density central
part. For certain observables, this {}``background'' contribution
completely spoils the {}``signal'', and to properly interpret the
data, we need to subtract this background.

In order to get quantitative results, we employ exactly the same method
as described recently \cite{corona} for analyzing RHIC heavy ion
data: we use EPOS \cite{epos}, which has proven to work very well
for pp and dAu collisions at RHIC and for pp scattering at SPS. \textbf{}EPOS
is a parton model, 

~\vspace*{7.8cm}

\noindent so in case of a nuclear collision there are many binary
interactions, creating partons, which then hadronize employing a phenomenological
procedure called string fragmentation. Here, we modify the procedure:
we have a look at the situation at an early proper time $\tau_{0}$,
long before the hadrons are formed: we distinguish between string
segments in dense areas (more than $\rho_{0}$ segments per unit proper
volume), from those in low density areas. We refer to high density
areas as core, and to low density areas as corona. In order to make
a quantitative statement, we adopt the following strategy: the low
density part will be treated using the usual EPOS particle production
which has proven to be very successful in pp and dAu scattering (the
peripheral interactions are essentially pp or pA scatterings). For
the high density part, we simply try to parameterize particle production,
in the most simple way possible. To do so, we consider the core contributions
sepatately in different longitudinal segments. Connected core regions
in a given segment are considered to be clusters, whose energy and
flavor content are complete determined by the corresponding string
segments. Clusters are then considered to be collectively expanding.
We assume that the clusters hadronize at some given energy density
$\varepsilon_{\mathrm{hadr}}$, having acquired at that moment a collective
radial flow, with a linear radial rapidity profile from inside to
outside, characterized by the maximal radial rapidity $y_{\mathrm{rad}}$.
In addition, we impose an azimuthal asymmetry proportional to the
initial space eccentricity. Hadronization then occurs according to
covariant phase space. For more details and parameters see \cite{corona}.

\textbf{So we employ exactly the same procedure as we did for RHIC,
with even the same parameters -- up to two exceptions, concerning
the parameters $\tau_{0}$ and $y_{\mathrm{rad}}$.} Whereas at RHIC
the final results are insensitive to variations of $\tau_{0}$ (in
the range 1-2fm), this is no longer the case at SPS, the reason being
the finite reaction time $\tau_{\mathrm{rea}}$ (time it takes for
the two nuclei to pass through each other), which is somewhat more
than a fermi. So we use $\tau_{0}=\tau_{\mathrm{rea}}$ , representing
the minimum possible value. \textbf{Since we have smaller initial
core densities at SPS compared to RHIC, we also expect a smaller radial
flow, so we take the freedom to use $y_{\mathrm{rad}}$ as a free
parameter, fixed by comparing to SPS data (the only free parameter
when going from RHIC to SPS!). We actually use $y_{\mathrm{rad}}=0.60$
at SPS, instead of $0.83$ at RHIC.}

\begin{figure}[tb]
\begin{center}\includegraphics[%
  scale=0.3,
  angle=270]{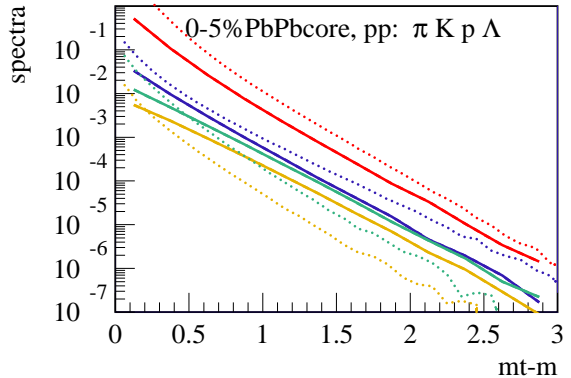}\end{center}
\vspace{-1.5cm}

\caption{Invariant yields $1/2\pi m_{t}\, dn/dydm_{t}$ of pions (red), kaons
(blue), protons (green), and lambdas (yellow), for the core contribution
corresponding of a central (0-5\%) PbPb collision (full lines) and
proton-proton scattering (dashed). The PbPb spectra are divided by
the number of collisions. \label{cap:core-pp}}
\end{figure}

\begin{figure}[tb]
\begin{center}\includegraphics[%
  scale=0.3,
  angle=270]{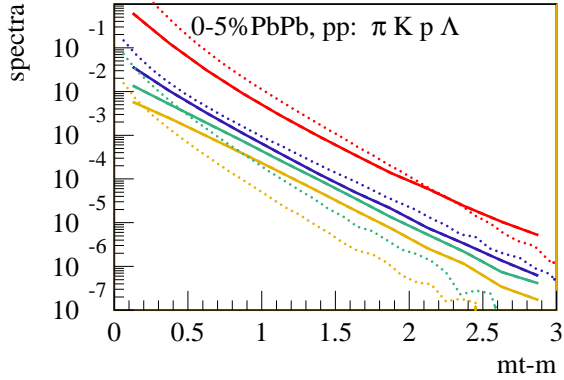}\end{center}
\vspace{-1.5cm}

\caption{Invariant yields $1/2\pi m_{t}\, dn/dydm_{t}$ of pions (red), kaons
(blue), protons (green), and lambdas (yellow), for the full contribution
(core + corona) of a central (0-5\%) PbPb collision (full lines) and
proton-proton scattering (dashed). The PbPb spectra are divided by
the number of collisions.\label{cap:all-pp}}
\end{figure}

\begin{figure}[tb]
\begin{center}\includegraphics[%
  scale=0.3,
  angle=270]{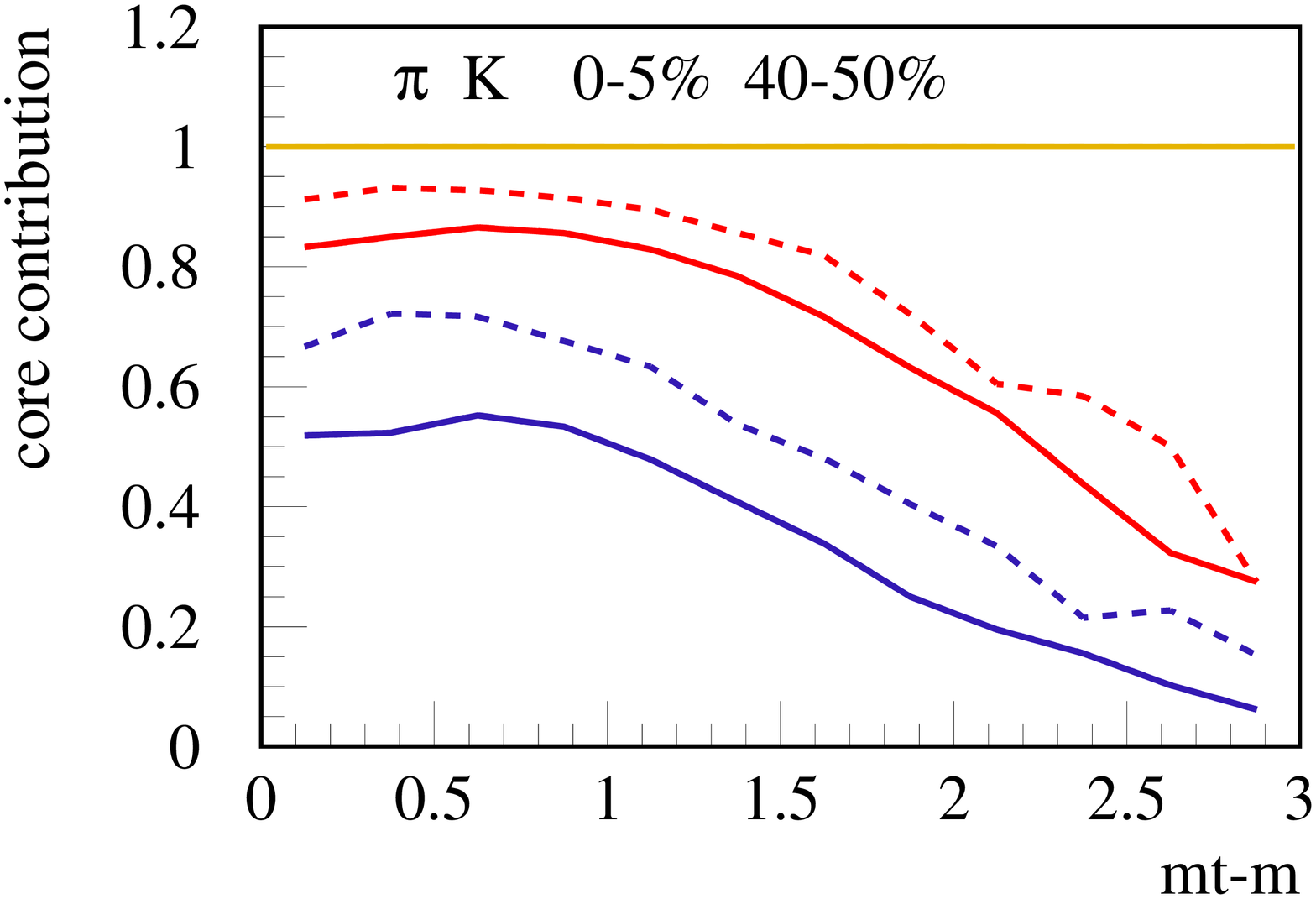}\end{center}
\vspace{-2cm}

\begin{center}\includegraphics[%
  scale=0.3,
  angle=270]{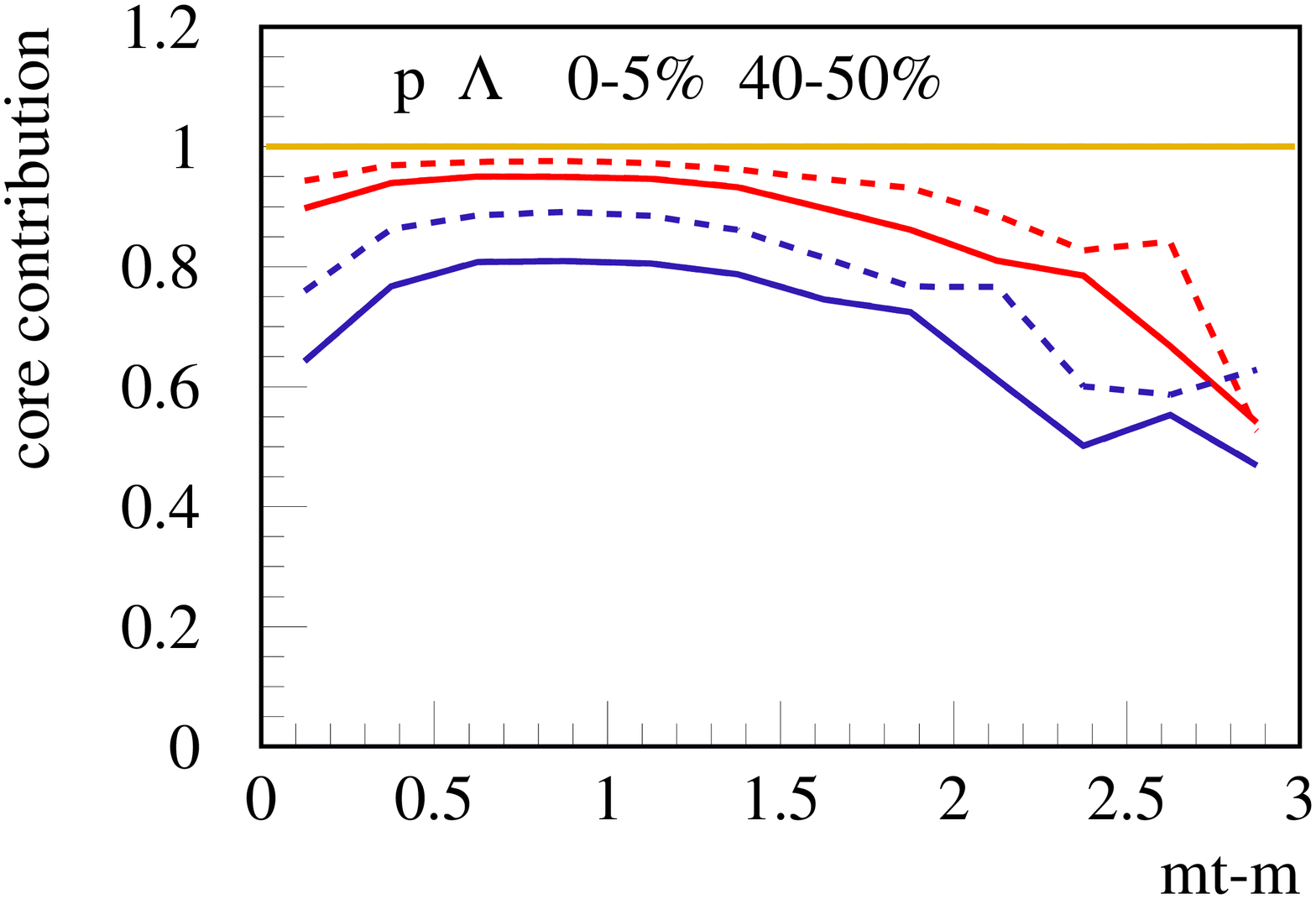}\end{center}
\vspace{-1.5cm}

\caption{The relative contribution of the core (core/(core+corona)) as a function
of the transverse mass for different centralities (0-5\%: red, 40-50\%:
blue. Upper figure: pions (full) and kaons (dashed). Lower figure:
protons (full) and lambdas (dashed).\label{cap:core-corona}}
\end{figure}

In the following, we will discuss results for PbPb collisions at 158
GeV. In fig. \ref{cap:core-pp}, we compare the core contribution
corresponding to a central (0-5\%) PbPb collision (which means purely
statistical hadronization, with flow) with pp scattering. We plot
$m_{t}$ spectra of pions, kaons, protons, and lambdas, the nuclear
spectra are divided by the number of binary collisions (according
to Glauber). Apart of the shape differences, the most striking feature
is the fact that the yields for the different pp contributions are
much wider spread than the core contributions, even more than at RHIC.
In pp (string fragmentation), lambdas (and even more the multi-strange
baryons) are very much suppressed compared to pion production, and
this suppression is even stronger at SPS compared to RHIC (since the
string masses are smaller).%
In fig. \ref{cap:all-pp}, we compare the full contribution (core
and corona) of a central (0-5\%) PbPb collision and proton-proton
scattering . The PbPb pion line crosses the pp line at large $p_{t}$,
which is due to the fact that the core contains particles from pA
like collisions, where we have a {}``Cronin-enhancement'' due to
parton ladder splitting, discussed in detail in \cite{epos}, which
has been introduced in order to understand dAu scattering at RHIC.%
In fig. \ref{cap:core-corona}, we plot the relative contribution
of the core (relative to the complete spectrum, core + corona) as
a function of $m_{t}-m$, for different particle species.%
For central collisions, the core contribution dominates largely (around
90\%), whereas for semi-central collisions (40-50\%) the core contribution
decreases, giving more and more space for the corona part. The precise
$m_{t}$ dependence of the relative weight of core versus corona depends
on the particle type.

\begin{figure}[tb]
\begin{center}\includegraphics[%
  scale=0.3,
  angle=270]{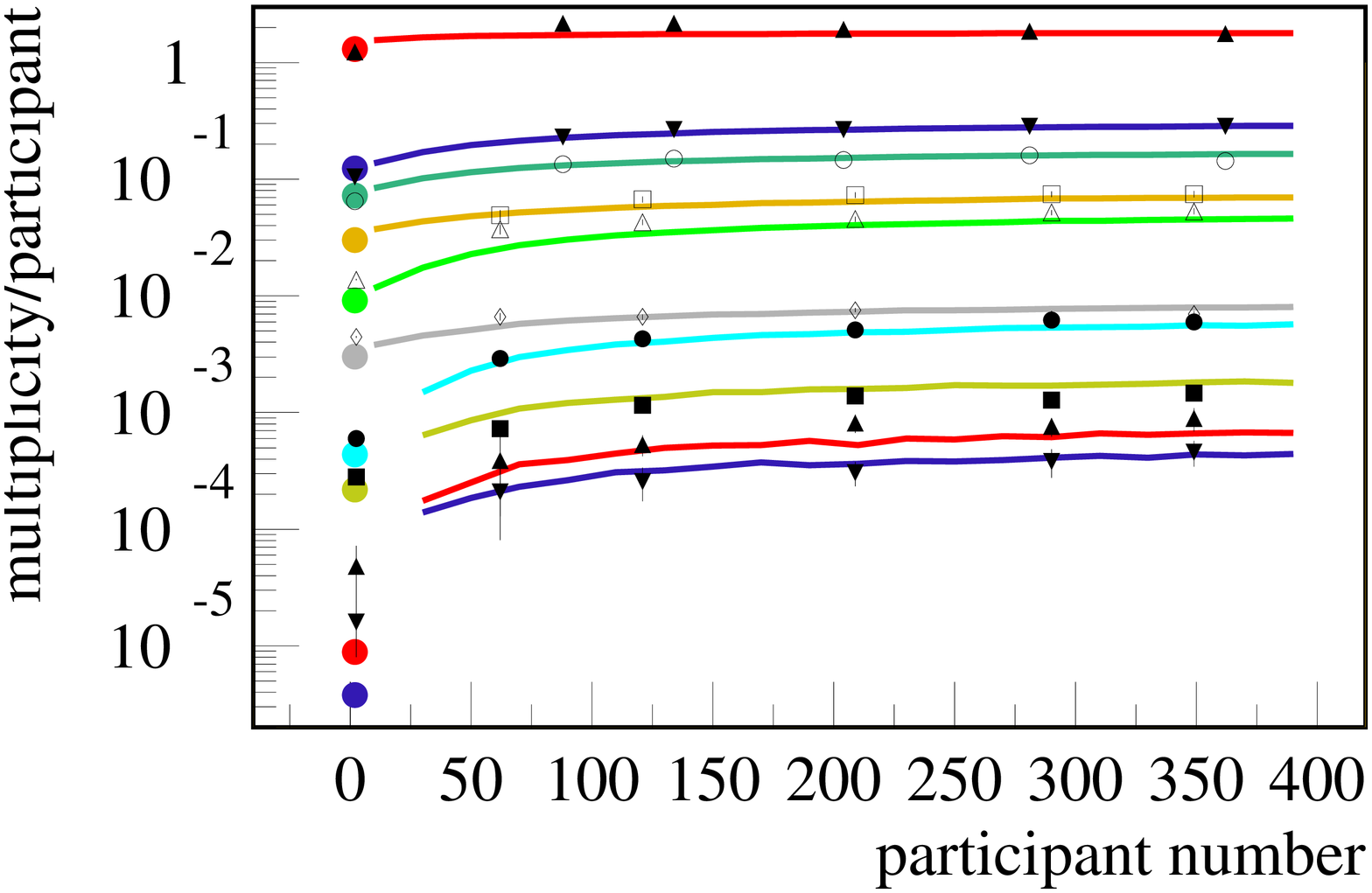}\end{center}
\vspace{-2cm}

\begin{center}\includegraphics[%
  scale=0.3,
  angle=270]{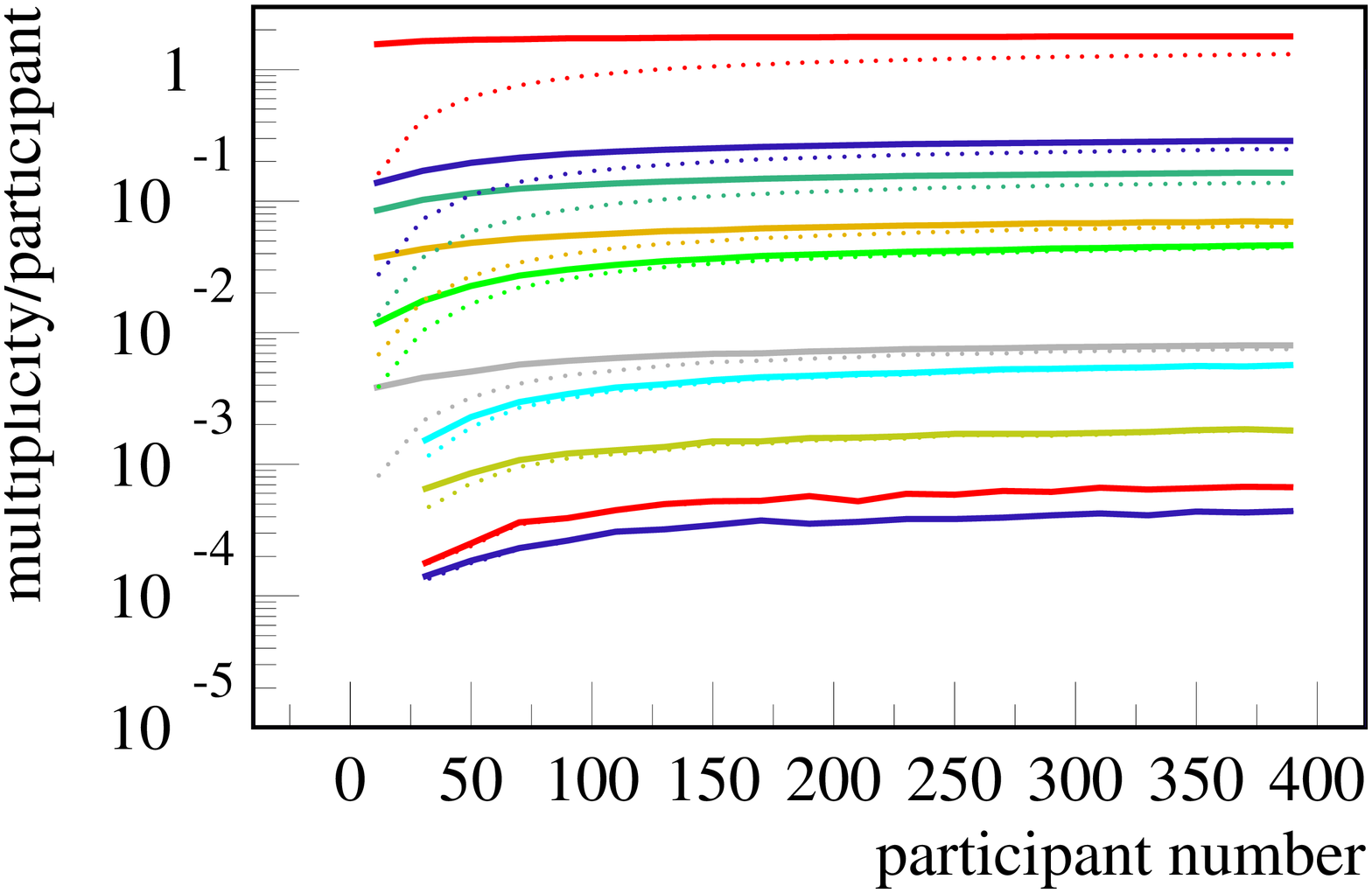}\end{center}
\vspace{-1.5cm}

\caption{\underbar{Upper diagram}: Particle yields per participant in PbPb
scattering at 158 GeV as a function of the number of participants,
for (from top to bottom): $\pi^{-}$, $K^{+}$, $K^{-}$, $K_{s}$,
$\Lambda$, $\bar{\Lambda}$, $\Xi$, $\bar{\Xi}$, $\Omega$, $\bar{\Omega}$.
We show data (points) \cite{na57centr} together with the full calculation
(core + corona, full line). The three upper-most curves refer to $4\pi$results,
all others to central rapidities. The leftmost points are pp calculations
(colored points) and pp or pBe data (black symbols).\label{cap:centrality}
\underbar{Lower diagram}: full calculations, same as in upper plot
(full lines) compared to the core contributions (dotted).}
\end{figure}

\begin{figure}[tb]
\begin{center}\includegraphics[%
  scale=0.3,
  angle=270]{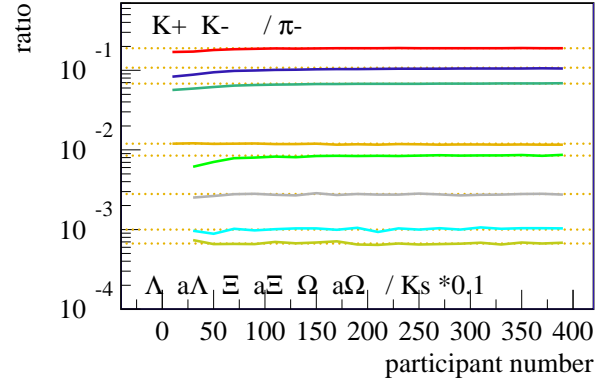}\end{center}
\vspace{-1.5cm}

\caption{Particle ratios as a function of centrality, from top to bottom:
$K^{+}/\pi^{-}$, $K^{-}/\pi^{-}$, $\Lambda/K_{s}\,\times$0.1 ,
$\bar{\Lambda}/K_{s}\:\times$0.1, $\Xi/K_{s}\:\times$0.1, $\bar{\Xi}/K_{s}\:\times$0.1,
$\Omega/K_{s}\:\times$0.1, $\bar{\Omega}/K_{s}\:\times$0.1. The
full lines are the calculations for the core contribution alone, the
dashed lines are horizontal lines, just to guide the eye. \label{cap:ratios}}
\end{figure}

We are now going to study SPS PbPb data (158 GeV). In fig. \ref{cap:centrality},
we plot the centrality dependence of the particle yield per participant
(per unit of rapidity), for different particle species, the data \cite{na49centr,na57centr}
together with the full calculation (upper diagram), and the full calculation
compared to the core contribution (lower diagram). %
The complete calculation follows quite closely the data. Whereas central
collisions are clearly core dominated, the core contributes less and
less with decreasing centrality. In fig. \ref{cap:ratios}, we consider
the corresponding particle ratios, as a function of centrality, for
the core contributions. We show the ratios of different particles,
with respect to $\pi^{-}$or $K_{s}$. The rations are practically
flat, apart of some decrease for very small participant numbers.%

\textbf{So our first important conclusion: after subtracting the {}``corona
background'', the interesting part, the core contribution, shows
an extremely simple behavior: there is no centrality dependence, the
systems are simply changing in size. The participant number is certainly
not a good measure of the volume of the core part, this is why the
overall multiplicities per participant decrease with decreasing centrality.
This is the same conclusion as in \cite{corona} for AuAu at RHIC.
And not only the SPS core contribution is as simple as the RHIC one,
it is even parameterized with the same parameters, apart of somewhat
more flow at RHIC!}

What makes the measured centrality dependence look complicated is
the mix of core and corona, depending on the particle species. Actually
the $\Omega$ are the simplest, since here the corona contribution
is negligible. Here, all the participants which contribute to the
corona (being more and more frequent with decreasing centrality) do
not contribute at all to the $\Omega$ production. This is why the
yields drop so strongly at small participant numbers. This also reflects
the fact that the core volume is not proportional to the number of
participants.

\begin{figure}[tb]
\vspace{-1.2cm}
\begin{center}\includegraphics[%
  scale=0.3,
  angle=270]{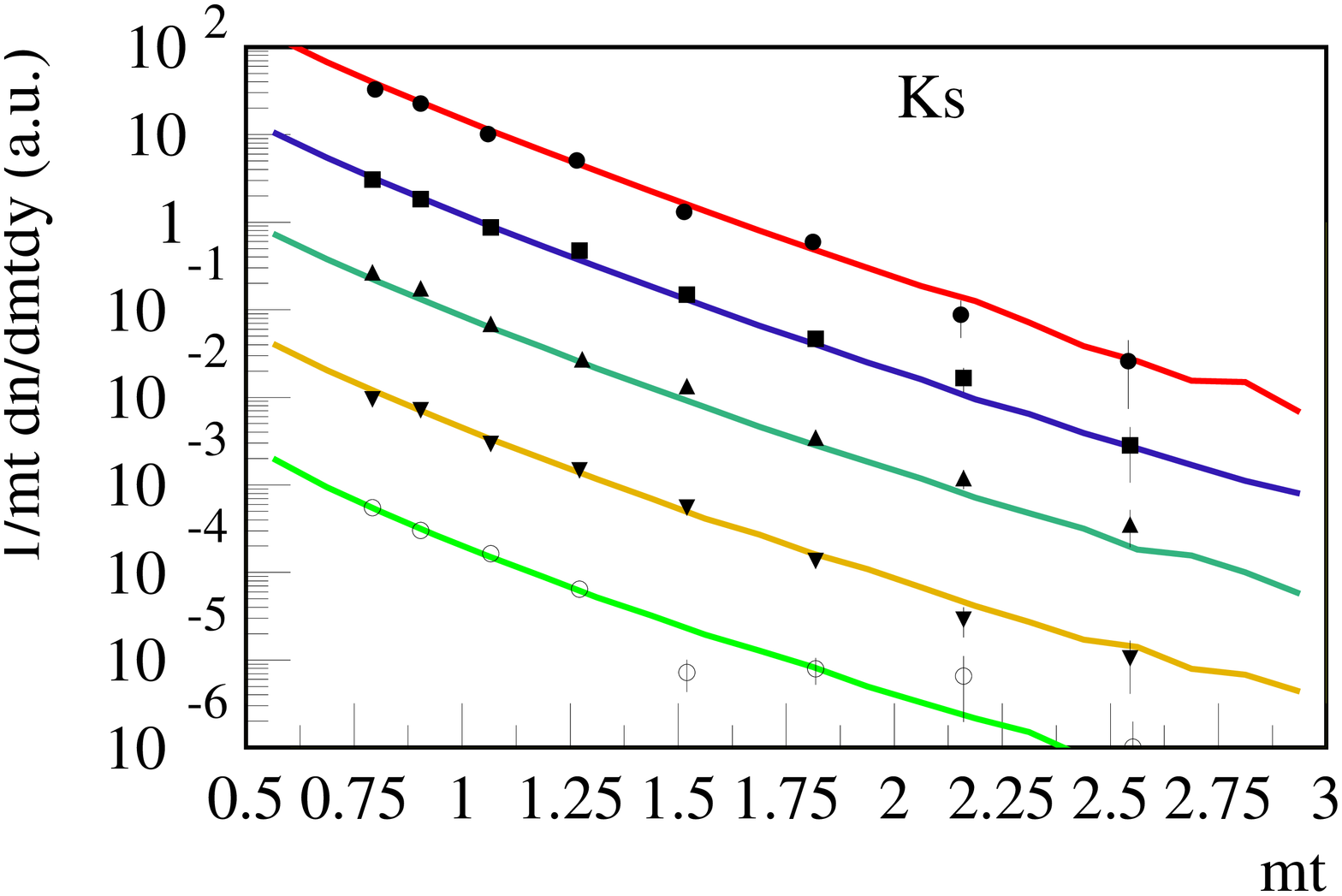}\end{center}

\vspace{-2.5cm}
\begin{center}\includegraphics[%
  scale=0.3,
  angle=270]{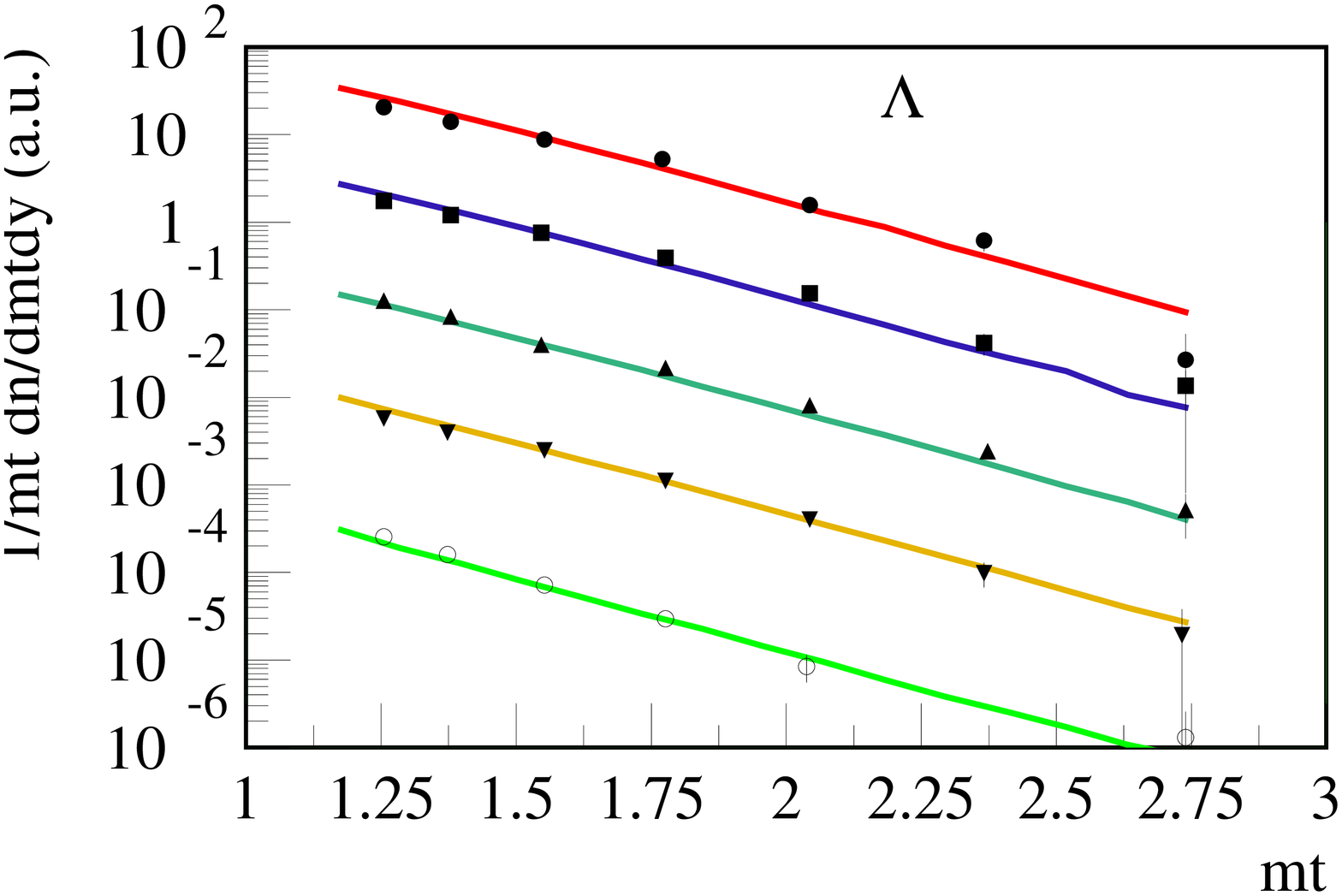}\end{center}
\vspace{-2.5cm}

\begin{center}\includegraphics[%
  scale=0.3,
  angle=270]{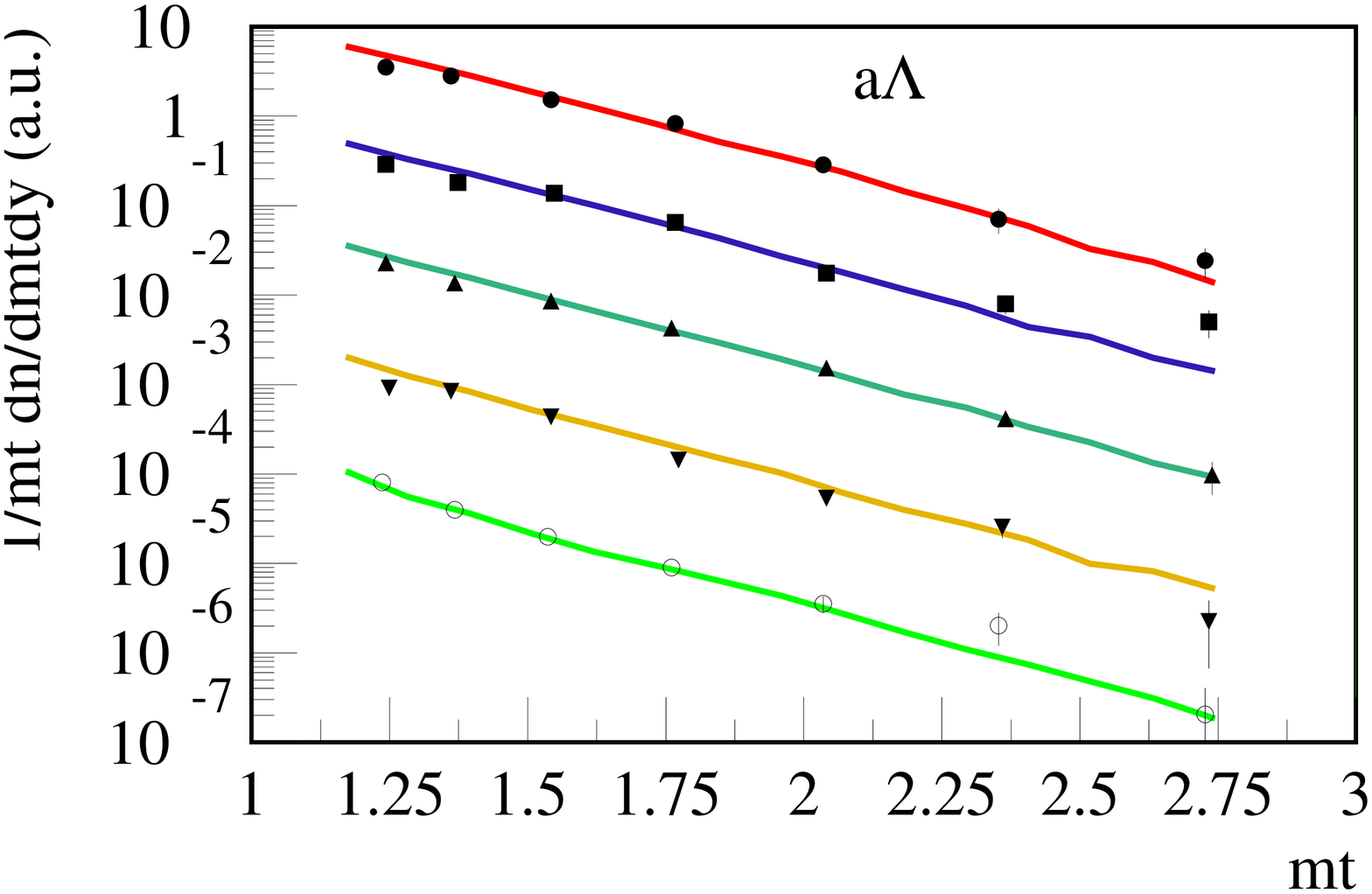}\end{center}
\vspace{-1.5cm}

\caption{Transverse mass spectra of $K_{s}$ (upper plot), $\Lambda$ (middle),
$\bar{\Lambda}$ (lower) in PbPb scattering at different centralities.
From top to bottom: 0-5\%, 5-11\%, 11-23\%, 23-40\%, 40-53\%. Lines
are full calculations, points are data \cite{na57mt}. The different
curves are displaced by factors of 10.\label{cap:kas}}\vspace{-0.3cm}

\end{figure}

\begin{figure}[tb]
\vspace{-1.2cm}
\begin{center}\includegraphics[%
  scale=0.3,
  angle=270]{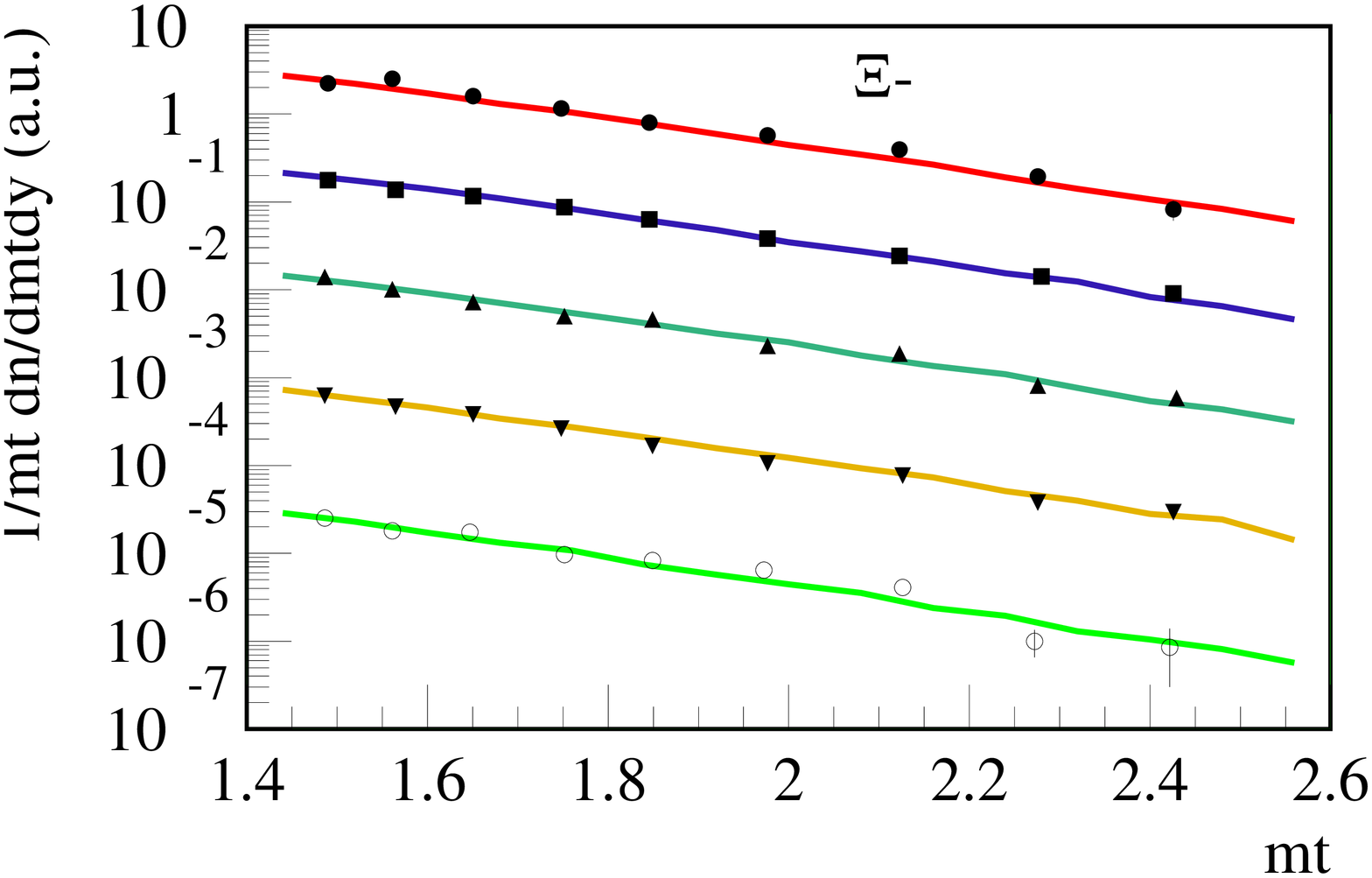}\end{center}

\vspace{-2.5cm}
\begin{center}\includegraphics[%
  scale=0.3,
  angle=270]{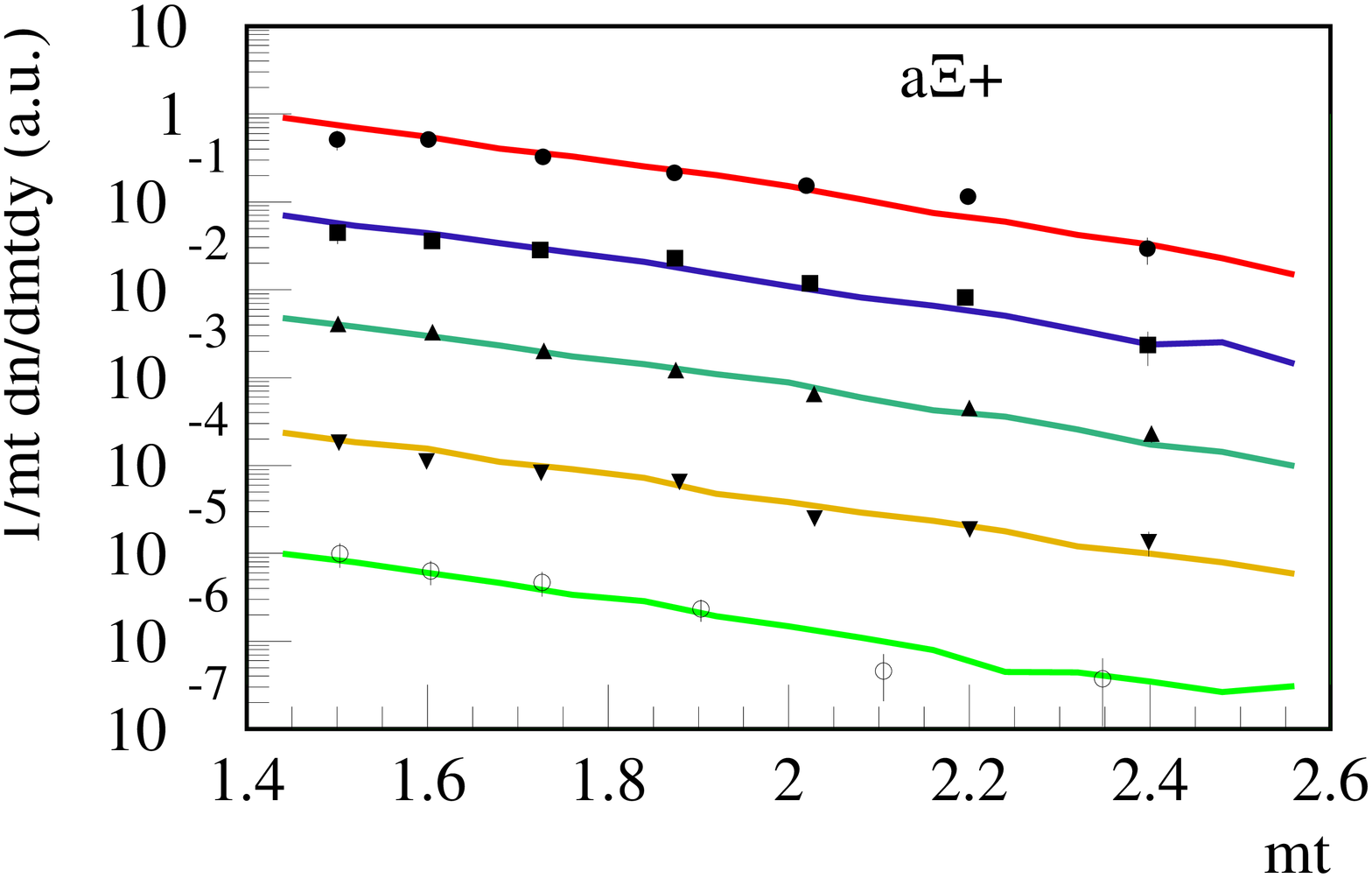}\end{center}
\vspace{-2.5cm}

\begin{center}\includegraphics[%
  scale=0.3,
  angle=270]{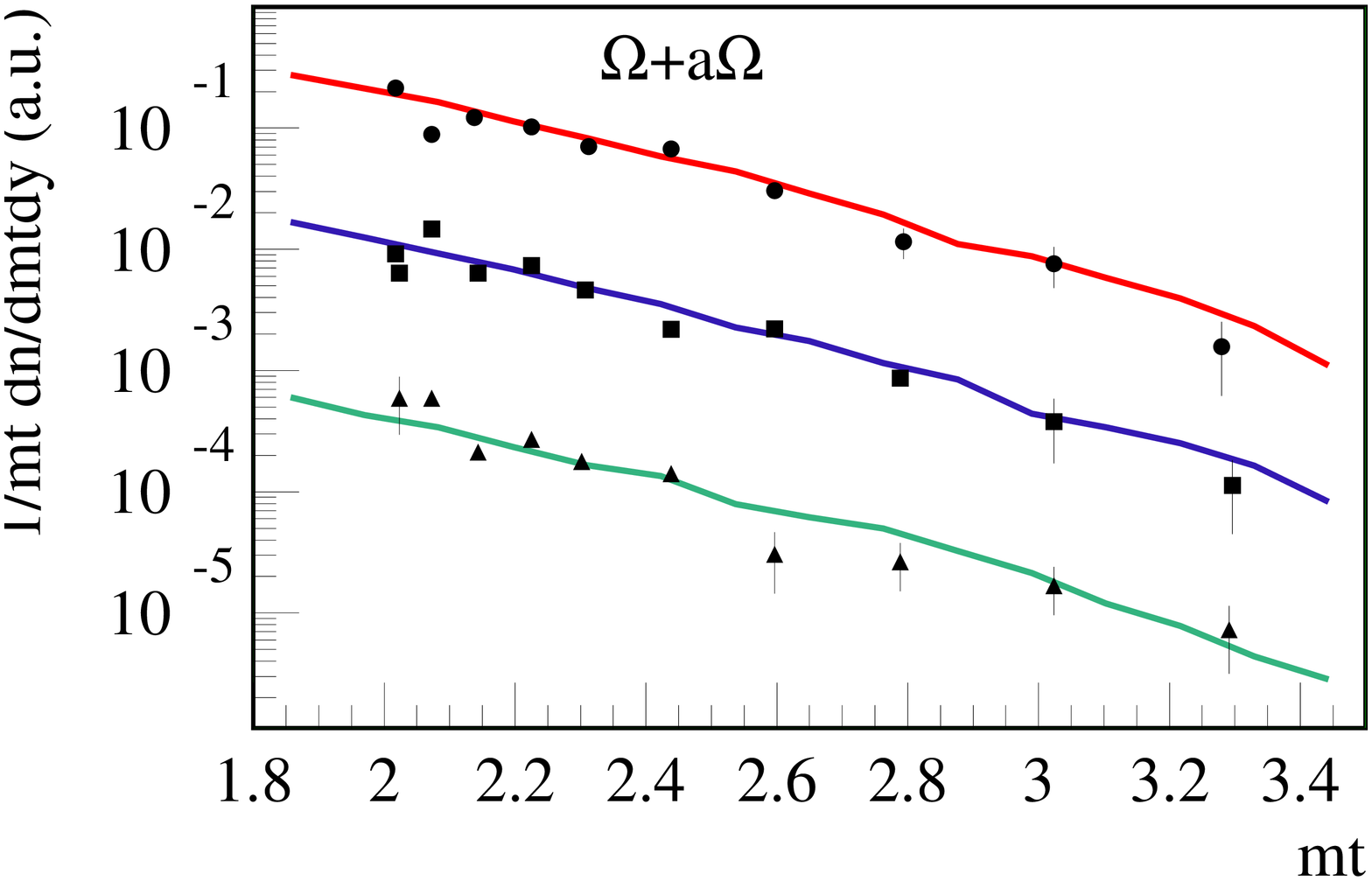}\end{center}
\vspace{-1.5cm}

\caption{Transverse mass spectra of $\Xi$ (upper plot), $\bar{\Xi}$ (middle),
$\Omega+\bar{\Omega}$ (lower) in PbPb scattering at different centralities.
From top to bottom: 0-5\%, 5-11\%, 11-23\%, 23-40\%, 40-53\%. Lines
are full calculations, points are data \cite{na57mt}. The different
curves are displaced by factors of 10.\label{cap:xim}}\vspace{-0.3cm}

\end{figure}

To demonstrate consistency, we check in the following $m_{t}$ spectra
for different hadron species, for PbPb collisions at 158 GeV, see
figs. \ref{cap:kas} and \ref{cap:xim}.%
The calculations represent the full contribution, core plus corona,
whose relative%
 contributions (at least for kaons and lambdas) can be obtained from
fig. \ref{cap:core-corona}. So it is actually a non-trivial superposition
of these two contributions, hiding the shape of the interesting part,
the core.

It should be noted that NA49 \cite{na49lda} reports much less lambda
production at central rapidity than NA57, and correspondingly our
calculation of $dn/dy(y=0)$ (being close to the NA57 values) is about
40\% above the NA49 results (whereas the transverse mass spectra almost
agree!). 

\begin{figure}[tb]
\vspace{-1.2cm}
\begin{center}\includegraphics[%
  scale=0.3,
  angle=270]{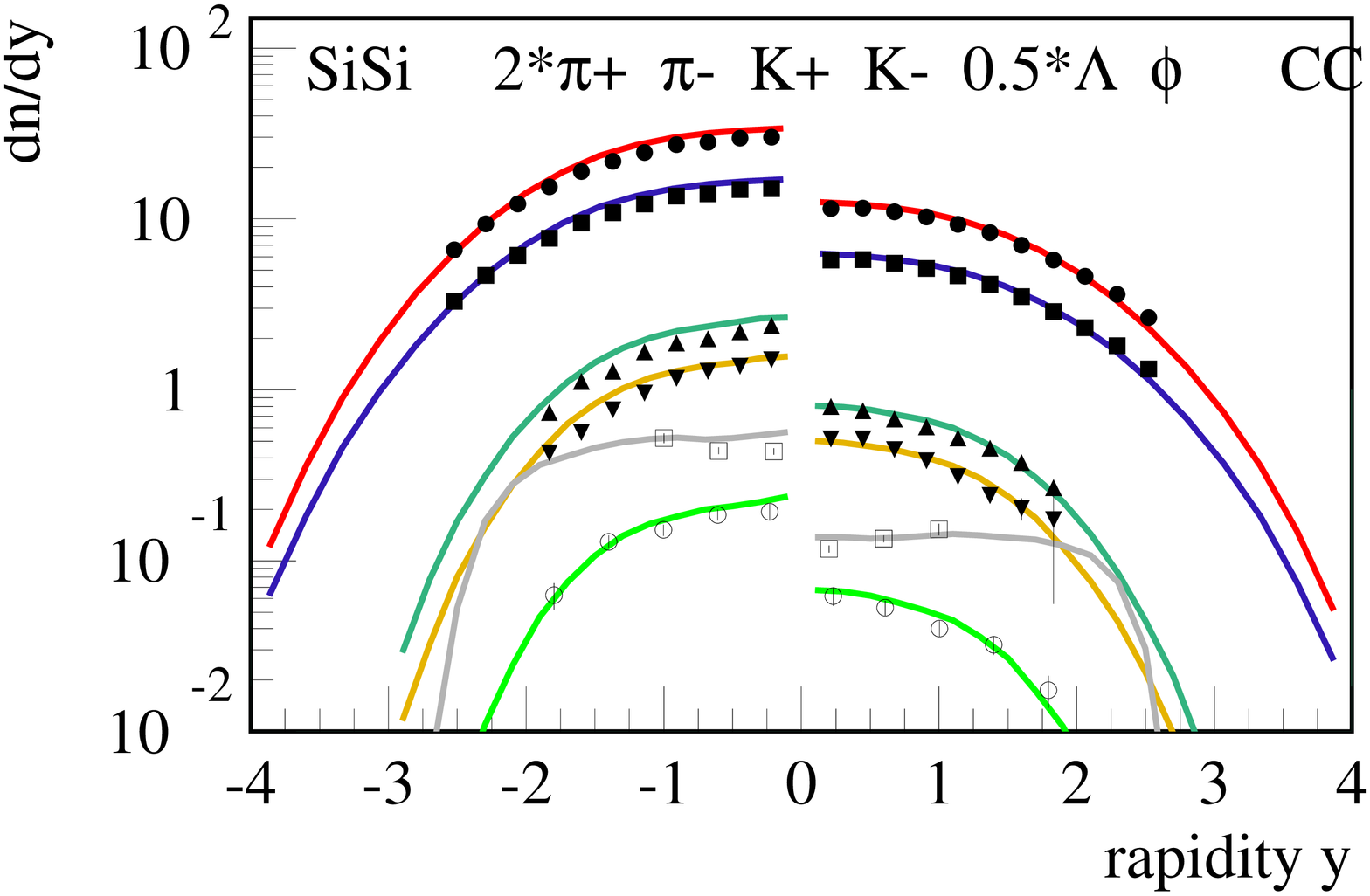}\end{center}

\vspace{-2.5cm}
\begin{center}\includegraphics[%
  scale=0.3,
  angle=270]{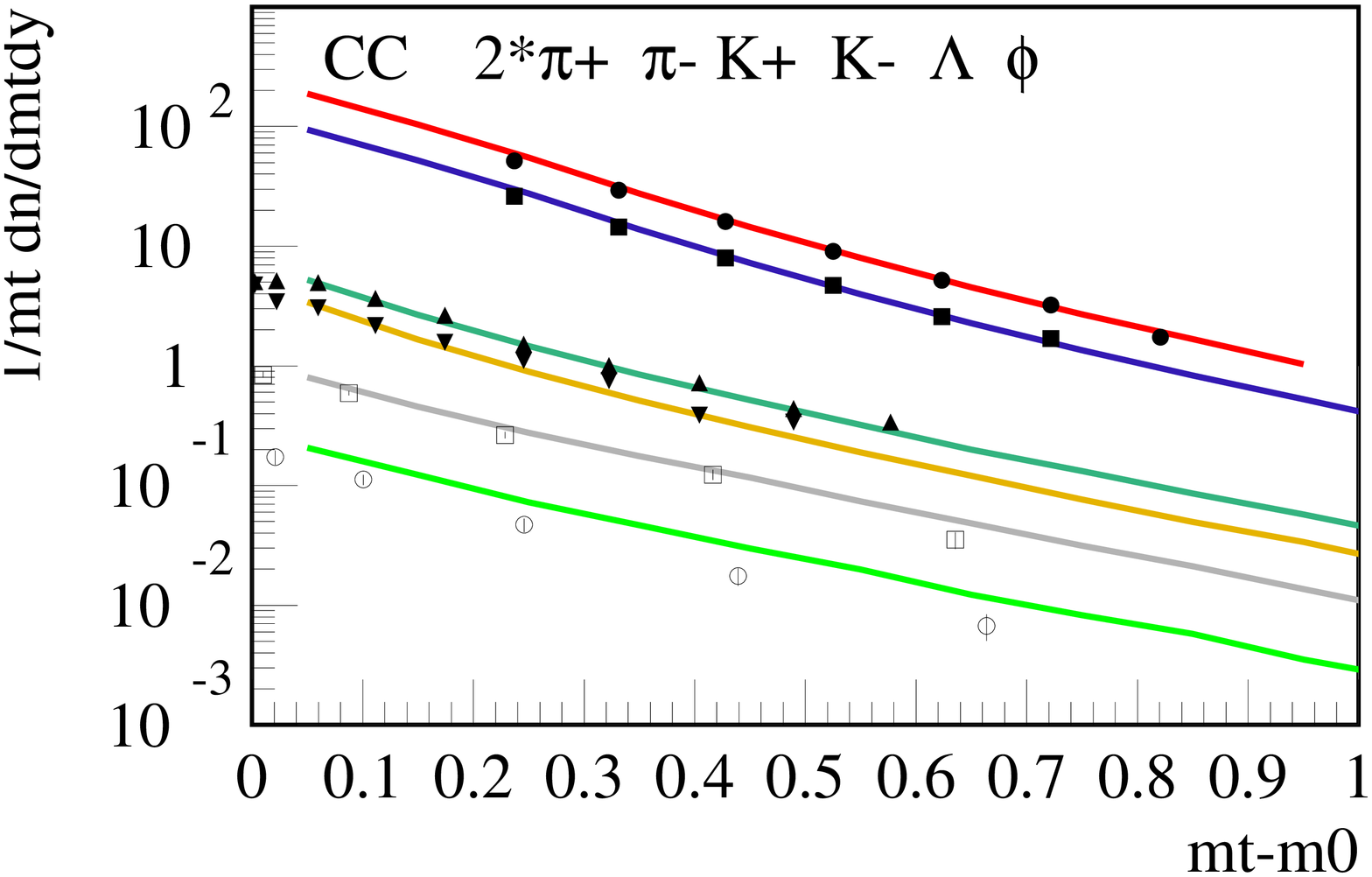}\end{center}
\vspace{-2.5cm}

\begin{center}\includegraphics[%
  scale=0.3,
  angle=270]{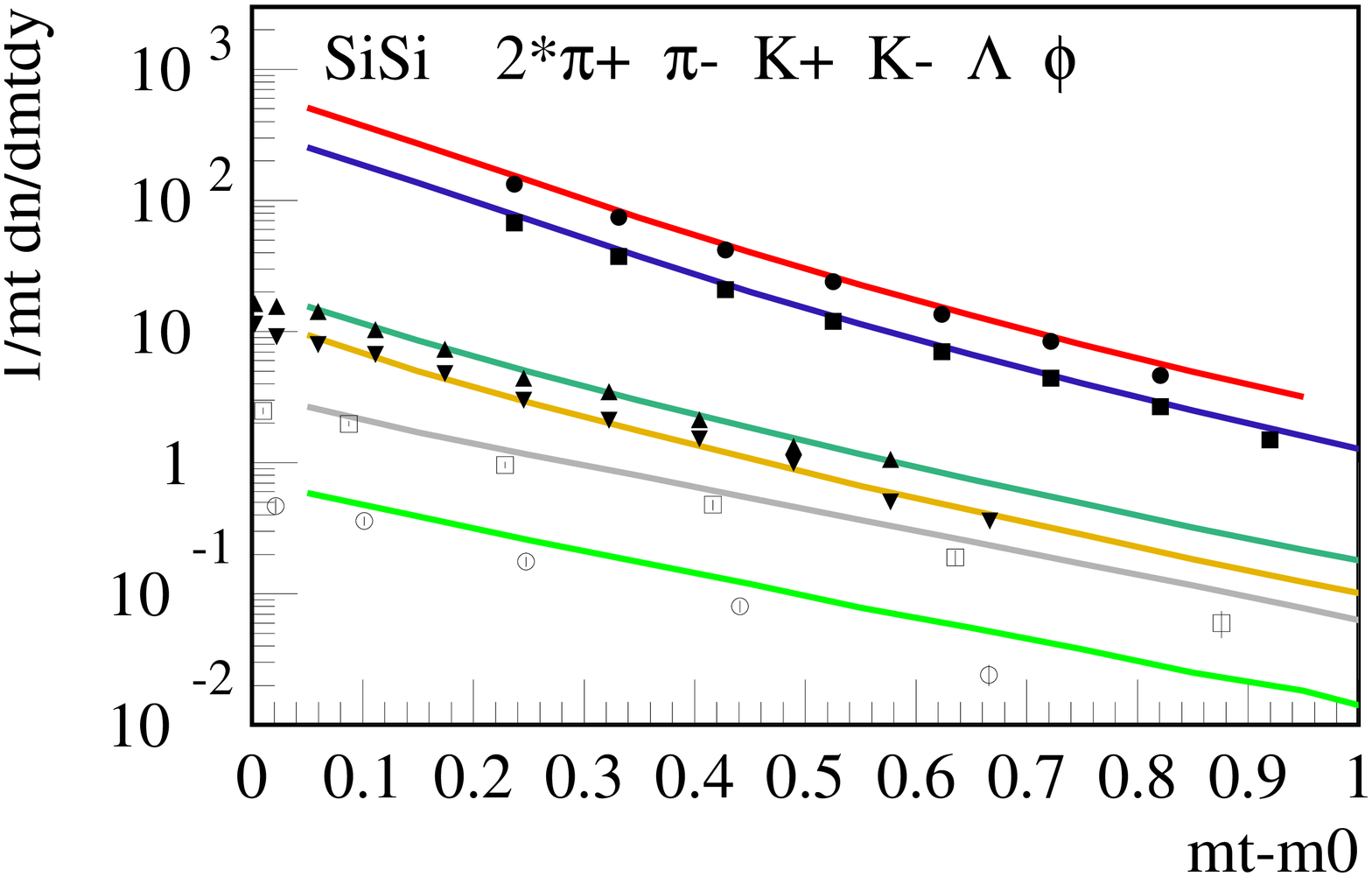}\end{center}
\vspace{-1.5cm}

\caption{Rapidity spectra of identified hadrons in central SiSi (upper, left)
and central CC (upper right) and transverse mass spectra in central
CC (middle) and central SiSi (bottom). In each figure, from top to
bottom: $\pi^{+}$, $\pi^{-}$, $K^{+}$, $K^{-}$, $\Lambda$, $\phi$.
Lines are full calculations, points are data. \label{cap:cc}}\vspace{-0.3cm}

\end{figure}

In fig. \ref{cap:cc}, we show rapidity and transverse mass spectra
of different hadron species for central CC and SiSi at 158 GeV. Again
we just show the full calculations (core plus corona) together with
the data, and we find a similar good agreement than in case of PbPb
(using the same parameters!), with the exception of the $\phi$ transverse
momentum spectrum, which is somewhat harder in the calculations. Concerning
the relative weights of core and corona, CC and SiSi is very similar
to peripheral PbPb, in the sense that the corona is relatively more
important than in central PbPb (the surface effect is bigger in small
nuclei than in big ones). 

To summarize: we have discussed the influence of the corona contribution
(occurring in the periphery of nuclear collisions) in PbPb, SiSi,
and CC collisions at the SPS. We provide a realistic treatment of
the corona, by using a model which works excellently for pp and pA.
We can provide a parametrization of the core, such that the complete
calculation (core + corona) provides a good fit of the observed particle
spectra in heavy ion collisions at the SPS. Our core results represent
to some extent {}``background corrected results'', removing the
undesired corona contributions from the spectra. \textbf{The core
shows an extremely simple centrality and species dependence: actually
there is no centrality dependence, all particle ratios are constant.
In other words, the cores in all these different systems (CC, SiSi,
PbPb at different centralities) are all identical, apart of a trivial
volume difference. Even more remarkable: the SPS results (concerning
core) are identical to the RHIC results, with one exception: there
is 30\% more radial flow at RHIC. So what really makes the experimental
results complicated is the non-trivial admixture of corona contribution
to the core.}

All the results concerning SPS in this paper have been obtained with
exactly the same procedure (and same parameters) as in a previous
study of RHIC results \cite{corona}, with just one exception: the
radial flow parameter. All the centrality and system size dependence
is purely determined by geometry.

\end{document}